\documentclass[reqno,12pt]{amsart}
\usepackage{amsmath, latexsym, amsfonts, amssymb, amsthm, amscd}
\usepackage{epsfig,epsf,psfrag}

\advance\hoffset -.75cm

\numberwithin{equation}{section}

\newtheorem{theo}{Theorem}[section]

\newtheorem{rem}[theo]{Remark}

\DeclareMathSymbol{\leqslant}{\mathalpha}{AMSa}{"36} 
\DeclareMathSymbol{\geqslant}{\mathalpha}{AMSa}{"3E} 
\DeclareMathSymbol{\eset}{\mathalpha}{AMSb}{"3F}     
\renewcommand{\leq}{\;\leqslant\;}                   
\renewcommand{\geq}{\;\geqslant\;}                   

\newcommand{\bra}{\langle}
\newcommand{\ket}{\rangle}

\newcommand{\cE}{\ensuremath{\mathcal E}}
\newcommand{\cF}{\ensuremath{\mathcal F}}
\newcommand{\cG}{\ensuremath{\mathcal G}}

\newcommand{\cI}{\ensuremath{\mathcal I}}

\newcommand{\cL}{\ensuremath{\mathcal L}}

\newcommand{\bbA}{{\ensuremath{\mathbb A}} }

\newcommand{\bbC}{{\ensuremath{\mathbb C}} }

\newcommand{\bbE}{{\ensuremath{\mathbb E}} }

\newcommand{\bbP}{{\ensuremath{\mathbb P}} }

\newcommand{\ga}{\alpha}
\newcommand{\gb}{\beta}
\newcommand{\gga}{\gamma}            

\newcommand{\gep}{\varepsilon}       
\newcommand{\gp}{\varphi}
\newcommand{\gr}{\rho}

\newcommand{\gP}{\Phi}

\newcommand{\gl}{\lambda}

\newcommand{\gs}{\sigma}

\newcommand{\dQ}{{\dot Q}}
\newcommand{\dK}{{\dot K}}

\title{Current large deviations in a driven dissipative model}

\author{T. Bodineau}
\address{Ecole Normale Sup\'erieure, DMA, 45 rue d'Ulm
75230 Paris cedex 05, France}
\author{M. Lagouge}
\date{\today}
\thanks{We are grateful to B. Derrida and J. Lebowitz for very useful discussions.
TB acknowledges the support of the French Ministry of Education through the ANR BLAN07-2184264 grant.
This work was partially supported by the NSF Grant DMR-044-2066 and AFOSR Grant AF-FA9550-04 during a stay at Rutgers University and by  the {\it Florence Gould Foundation Endowment} during a stay at the Institute for Advanced Study.}

\begin{document}

\maketitle

\begin{abstract}
We consider lattice gas diffusive dynamics with creation-annihilation in the bulk and maintained out of equilibrium by two reservoirs at the boundaries.
This stochastic particle system can be viewed as a toy model for granular gases where the energy is injected at the boundary and dissipated in the bulk. 
The large deviation functional for the particle currents flowing through the system is computed and some 
physical consequences are discussed: the mechanism for local current fluctuations, dynamical phase transitions, the fluctuation-relation.
\end{abstract}

\section{Introduction}

In different contexts like biological systems or chemical reactions, many out of equilibrium dynamics combine diffusive and dissipative features which can be described macroscopically by reaction-diffusion equations. 
In order to analyze these systems from a microscopic point of view, a lattice gas model with fast stirring and a local creation-annihilation mechanism has been introduced 
in \cite{DFL} by combining a symmetric simple exclusion process (SSEP) and a Glauber dynamics.
Another example of driven dissipative system is a granular gas where the energy is dissipated through inelastic collisions. Direct analytic approaches of the granular gases are notoriously difficult, but stochastic toy models have been proposed in order to capture some aspects of the physics \cite{B,LL,SL,F,FP} or for simulation purposes \cite{GRW}. 
The particle system introduced in \cite{DFL} in contact with two heat bath shares also similar features.

\medskip

In non-equilibrium statistical mechanics, there is no analogous to the Gibbs formalism and the large deviation functional can be viewed as a substitute for the free energy.
In particular, the large deviations of the current received a lot of attention over the last few years (see \cite{BDGJL6,D,G} for reviews). Exact expressions of the large deviation functional have been obtained for conservative diffusive stochastic dynamics in contact with reservoirs \cite{BDGJL4}--\cite{BDL}. 
In this paper, we generalize these results to the non-conservative stochastic dynamics which combine a fast stirring and a Glauber dynamics as in \cite{DFL}. The relevant macroscopic parameters to describe these stochastic systems are the density, the conservative current (from the stirring dynamics) and the non-conservative current (from the Glauber dynamics).
The large deviation functional \eqref{eq: LDF} gives the exponential cost of observing a deviation of these 
three parameters simultanously.
The large deviations for the density have been already studied in \cite{JLV} and the long-range correlations of the steady state in \cite{BJ}. 
The large deviations of the injected power were computed for a different non-conservative stochastic dynamics in \cite{FP}.

\medskip

The large deviation functional derived in \eqref{eq: LDF}  measures the cost of observing a joint deviation of the density and the currents. 
Thus the density large deviations obtained in \cite{JLV} can be recovered from the functional \eqref{eq: functional} by optimizing the deviations on the currents. In the same way, observing an atypical conservative current imposes a non trivial constraint on the density and the non-conservative current. 
As a consequence, we argue in section \ref{sec: variance} that the local conservative current fluctuations are produced by a different mechanism than in the case of conservative dynamics.

For some choice of the Glauber rates, the steady state may exhibit a phase transition. This instability of the steady state leads to a new kind of dynamical phase transition for the current which is  presented in section \ref{sec: phase transition}.

The fluctuation relation for the entropy production has been investigated in the framework of dissipative systems in \cite{BGGZ,FP}. 
Using the exact expression of the large deviation functional \eqref{eq: functional}, we discuss its symmetry properties with respect to time reversal and recover the fluctuation relation for the entropy production. Contrary to the conservative dynamics, the entropy production is no longer proportional to the current injected in the system, but depends in a complex way on the currents and the density.

\section{Models and Notation}

To fix the idea, we first recall in section \ref{subsec: microscopic} the microscopic model introduced in \cite{DFL}. 
As it will be clear from section \ref{sec: hydrodynamic limit}, the details of the microscopic evolution will not be important to study the macroscopic limit and one can consider more general dynamics than the one introduced in section \ref{subsec: microscopic}. The relevant macroscopic parameters are defined in section \ref{sec: hydrodynamic limit} and some concrete examples are then discussed in section \ref{sec: Some concrete examples}.

\subsection{A microscopic model}
\label{subsec: microscopic}

We consider a particle system on a one dimensional chain with $N$ sites $\{1,N\}$.
The state of each site $i$ can be occupied or empty and is encoded by $\eta_i \in \{0,1\}$.
During the infinitesimal time interval $dt$, each particle has a probability $N^2 dt$ of jumping to the left  if the left
neighboring site is empty, $N^2 dt$ of jumping to the right  if the right neighboring site is empty.
Furthermore at each site $i$, a creation or an annihilation can occur with probability $c(i,\eta) dt$, where 
the rate $c(i,\eta)$ can depend on the local configuration around the site $i$. The simplest rates are constant creation and annihilation rates $c^+, c^-$ in which case  $c(i,\eta) = c^+ (1 -\eta_i) + c^- \eta_i$. We will also consider more complicated rates leading to dynamical phase transitions \eqref{eq: rate transition}.
Finally, to model the effect of reservoirs of particles at the boundaries another creation/annihilation mechanism is acting on the boundary sites: during each time interval $dt$ a particle is created at site
1 (if it is empty) with probability $N^2 \ga_1^+ dt$ and removed with probability $N^2 \ga_1^- dt$ (if it is occupied). Similarly at site $N$, particles are created or removed at rate $N^2 \ga_N^\pm dt$.

We stress the fact that the stirring of the particles and the action of the reservoirs is much faster (by a factor $N^2$) than the creation/annihilation mechanism in the bulk. The particles perform random 
walks thanks to the stirring mechanism. Thus in the large $N$ limit,  a particle created at the boundary has a positive probability to go through the system without being annihilated. 
This is the correct scaling to ensure a competition between the two mechanisms of the dynamics and therefore
 a non trivial thermodynamic limit ($N \to \infty$).

\bigskip

Rephrased in mathematical terms, the previous dynamics is a Markov chain with generator given by
\begin{eqnarray*}
\cL f (\eta) &=&  N^2  \sum_{i \in \{1,N-1\}} \Big[ f(\eta^{i,i+1}) - f (\eta) \Big] + \sum_{i\in \{1,N\}}
c(i, \eta) \big( f(\eta^i) - f (\eta) \big)\\
&& \qquad \qquad + N^2 \sum_{i =1 \ {\rm or } \ N} (\ga_i^+ (1 -\eta_i) + \ga_i^- \eta_i) \big( f(\eta^i) - f (\eta) \big) \, ,
\end{eqnarray*}
where the configuration $\eta^{i,i+1}$ is obtained from $\eta$ by swapping the occupation numbers of the sites $i,i+1$ and 
the configuration $\eta^i$ is obtained from $\eta$ by changing the occupation number at site $i$ from $\eta_i$ to $1-\eta_i$.

\bigskip

The microscopic currents associated to the previous dynamics are defined as:

\noindent
{\it The conservative current.}
For a given edge $(i,i+1)$, let $Q_{T,(i,i+1)}$ be the number of jumps from site $i$ to site $i+1$ minus the number of jumps from site $i+1$ to site $i$ during the time interval $[0,T]$. 

\noindent
{\it The non-conservative current.}
We denote by $K^+_{T,i}$ the number of particles created at site $i$ during the time interval $[0,T]$. In the same way, $K^-_{T,i}$ denotes the number of particles annihilated at $i$. 

\medskip

Microscopically this leads to the following conservation law at each site $i$ and any time $t$
\begin{eqnarray}
\label{eq: micro conservation}
\eta_{t,i} - \eta_{0,i}  = K^+_{t,i} - K^-_{t,i} + Q_{t,(i-1,i)} - Q_{t,(i,i+1)} \, .
\end{eqnarray}

\subsection{The hydrodynamic limit}
\label{sec: hydrodynamic limit}

One can also  consider more general stirring processes than the SSEP of section \ref{subsec: microscopic} (e.g. the Zero range process or a Kawasaki dynamics \cite{spohn}). In the macroscopic limit ($N \to \infty$) this will lead to a non-trivial hydrodynamic limit provided that the diffusive dynamics is speeded up by a factor $N^2$ compared to the Glauber part. The details of the microscopic evolution will 
be averaged out in the thermodynamic limit and the macroscopic equations will depend only on a few relevant macroscopic parameters which we describe below.

\medskip

We first consider the {\it purely diffusive} evolution in contact with the reservoirs at the boundary (i.e. when $c(i,\eta)=0$).
According to the hydrodynamic formalism \cite{spohn}, a given
diffusive lattice gas can be characterized by the diffusion coefficient $D(\rho)$
and the conductivity  $\sigma(\rho)$ both depending on the density $\rho$.
Following  \cite{BD1}, one way to define these coefficients is to consider a one dimensional system of length $N$ connected to reservoirs at its two ends. Since the number of particles is preserved by the dynamics, it is equivalent to measure the current $Q_{T,(i,i+1)}$ through any bond $(i,i+1)$. 
When both reservoirs are at the same density $\rho$, the variance of the current $Q_{T,(i,i+1)}$  transfered during a long time $T$ from one reservoir to the other is given for large $N$ by
\begin{equation}
 \lim_{T \to \infty}  {\langle \big( Q_{T,(i,i+1)} \big)^2 \rangle \over T} = \sigma(\rho) N \, ,
\label{eq: sigma}
\end{equation}
where $\langle \cdot \rangle$ denotes the expectation of the dynamics starting from the invariant measure.
The conductivity $\sigma(\rho)$ is defined by \eqref{eq: sigma}.
Note that the scaling wrt $N$ comes from the fact that the diffusive dynamics has been speeded up with a rate $N^2$. 
On the other hand if the left reservoir is at density $\rho + \delta$ and the right reservoir at density $\rho$, the average current is given,  for small $\delta$ and large $N$, by
\begin{equation}
\lim_{T \to \infty} {\langle Q_{T,(i,i+1)} \rangle \over T} = D(\rho) N \delta \, ,
\label{eq: D}
\end{equation}
which is simply Fick's law and defines the function $D(\rho)$.
For the SSEP, $\sigma(\rho)=2 \rho(1-\rho)$ and $D(\rho)=1$ \cite{spohn}.

At time $t$ and for $x$ in $[0,1]$, the macroscopic density is denoted by $\gr(x,t)$ when $N \to \infty$
\begin{equation}
\gr(x,t) \simeq \frac{1}{2 \gep N} \sum_{|i - N x | \leq \gep} \eta_i(t) \, .
\label{eq: density}
\end{equation}
We stress the fact that the microscopic diffusion has been speeded up by a factor $N^2$ thus $t$ stands
already for a macroscopic time. 
At the boundary, the density is imposed by the reservoirs at any time $t$
\begin{equation}
\gr(0,t) = \gr_a = {\ga_1^+ \over \ga_1^+ + \ga_1^-}, 
\qquad  
\gr(1,t) = \gr_b= {\ga_N^+ \over \ga_N^+ + \ga_N^-} \, ,
\label{eq: boundary}
\end{equation}
where the reservoir densities are those of the dynamics defined in section \ref{subsec: microscopic}.

Finally we define the macroscopic conservative current at $x \in [0,1]$ up to time $t$
\begin{eqnarray}
\label{eq: total conservative}
Q(x,t)   \simeq \frac{1}{2 \gep N^2} \sum_{|i - N x | \leq \gep}  Q_{t,(i,i+1)} \, ,
\end{eqnarray}
where $\gep$ vanishes in the thermodynamic limit ($N \to \infty$).
Its time derivative $\dQ(x,t)$ stands for the local current at $x$ and according to \eqref{eq: D}, 
its typical value is given by
\begin{eqnarray}
\label{eq: mean conservative}
\dQ(x,t)  = - D(\gr(x,t)) \partial_x \gr(x,t) \, .
\end{eqnarray}
Thus the hydrodynamic limit is given by \cite{spohn}
\begin{equation}
\partial_t \gr(x,t)  = \partial_x \big( D(\gr(x,t)) \partial_x \gr(x,t) \big) \, ,
\label{eq: sans dissipation}
\end{equation}
with the boundary conditions \eqref{eq: boundary}.

\bigskip

We investigate now the hydrodynamic behavior of the {\it full dynamics}. Let $\mu_{\tilde \rho}$ be the invariant measure of the  stirring dynamics at the uniform density $\tilde \rho$.
(For the SSEP then $\mu_{\tilde \rho}$ is simply the Bernoulli product measure at density $\tilde \rho$).
At time $t=0$, the system starts at constant density $\tilde \rho$ from the measure $\mu_{\tilde \rho}$, then for any site $i = [N x]$ in the bulk
\begin{equation}
\partial_t \gr(x,t=0) = 
\partial_{t} \bra \eta_i (t=0) \ket =  - A ({\tilde \rho}) + C({\tilde \rho}) \, ,
\label{eq: dissipation 0}
\end{equation}
where the  macroscopic annihilation rate $A$ and the macroscopic creation rate $C$ 
are obtained by averaging the microscopic rates over the measure $\mu_{\tilde \rho}$
\begin{equation}
A ({\tilde \rho}) = \mu_{\tilde \rho} \big( c(i,\eta) \eta_i \big), \qquad
C ({\tilde \rho}) = \mu_{\tilde \rho} \big( c(i,\eta) (1-\eta_i) \big) \, .
\label{eq: AC}
\end{equation}
The measure $\mu_{\tilde \rho}$ is in general no longer invariant for the full dynamics, however the fast stirring
(with an $N^2$ rescaling) dominates locally and maintains the system in a local equilibrium. 
Even though, the density $\rho(x,t)$ evolves under the action of the dynamics, the local statistics remain 
given by $\mu_{\rho(x,t)}$ up to some small corrections which are vanishing for large $N$.
Thus when $N \to \infty$,  \eqref{eq: dissipation 0} can be generalized to later times by adding the diffusion term 
\eqref{eq: sans dissipation} and the density relaxes by following  the hydrodynamic equation
\begin{equation}
\partial_t \gr(x,t)  = \partial_x \big( D(\gr(x,t)) \partial_x \gr(x,t) \big) - V' \big( \gr(x,t) \big) \, ,
\label{eq: dissipation}
\end{equation}
where the potential is given by 
\begin{equation}
V' \big( u \big)  = A \big( u \big) -  C \big( u \big) \, ,
\label{eq: potential}
\end{equation}
and the boundary conditions are imposed by the left and right reservoirs $\gr(0,t) = \gr_a$ and $\gr(1,t)= \gr_b$ \eqref{eq: boundary}.
In particular, the steady state density $\bar \gr$ satisfies
\begin{equation}
\partial_x \big( D(\bar \gr(x)) \partial_x \bar \gr(x) \big) - V' \big( \bar \gr(x) \big) = 0  \, .
\label{eq: steady}
\end{equation}
These heuristic ideas have been rigorously justified in \cite{DFL,JLV,BL} for the microscopic model of section \ref{subsec: microscopic} (in this case $D=1$).

\medskip

The macroscopic non-conservative current at $x \in [0,1]$ up to time $t$ is defined by
\begin{eqnarray}
\label{eq: total non-conservative}
K(x,t)   \simeq \frac{1}{2 \gep N} \sum_{|i - N x | \leq \gep} K^+_{t,i} - K^-_{t,i}  \, .
\end{eqnarray}
Its time derivative $\dK(x,t)$ stands for the local current at $x$ and is typically given by
\begin{eqnarray}
\label{eq: K macro}
\dK(x,t)  = - A (\gr(x,t)) + C(\gr(x,t)) \, .
\end{eqnarray}

\medskip

After rescaling, the microscopic conservation law \eqref{eq: micro conservation} reads
\begin{eqnarray}
\label{eq: macro conservation}
\partial_t \rho(x,t) = - \partial_x \dQ(x,t) + \dK(x,t) \, .
\end{eqnarray}

\subsection{Some concrete examples}
\label{sec: Some concrete examples}

The previous models can be used to describe a wide variety of physical systems.
We describe below a few applications.

\medskip

\noindent
{\bf Granular gasses.}

When there is no creation in the bulk,
the dynamics introduced in section \ref{sec: hydrodynamic limit} can be viewed as  a toy model to mimick the granular gases where energy is injected at the boundary and dissipated in the bulk by inelastic collisions.
Similar dynamics have been considered in \cite{B,LL,SL}. More precisely, for the model of section \ref{subsec: microscopic}, if $c(i, \eta) = \alpha \eta_i$, then the hydrodynamic limit equation \eqref{eq: dissipation} reads
\begin{equation}
\label{eq: example granular}
\partial_t \gr(x,t)  = \partial^2_x  \gr(x,t)  - \alpha \gr(x,t) \, ,
\end{equation}
with boundary conditions $\gr_a,\gr_b$ \eqref{eq: boundary}.
$\rho (x,t)$ can be interpreted as a local energy dissipated at rate $\alpha \gr$ in the bulk
and the stationary state \eqref{eq: steady} satisfies  
\begin{equation*}
\bar \gr(x) = \gr_a \cosh ( \sqrt{\alpha} x) + {\gr_b - \gr_a  \cosh ( \sqrt{\alpha} ) \over \sinh(\sqrt{\alpha})} \sinh(\sqrt{\alpha} x)  \, .
\end{equation*}

\bigskip

\noindent
{\bf Reaction-diffusion.}

The microscopic model of section \ref{subsec: microscopic} was originally introduced \cite{DFL} to model diffusion reaction type equations often used to describe chemically reacting systems. 
For a Glauber dynamics which satisfies detailed balance with respect to the Gibbs measure of a one dimensional Ising model with nearest neighbor interaction at inverse temperature $\gb$, then 
\begin{eqnarray}
\label{eq: rate transition}
c(i,\eta) =1 + \gga (1 - 2 \eta_i) 2 ( \eta_{i-1} + \eta_{i+1}- 1)  + (-1 + 2 \eta_{i-1}) (-1 + 2 \eta_{i+1}) \gga^2 \, ,
\end{eqnarray}
with $\gga = \tanh(\gb)$. The hydrodynamic limit \eqref{eq: dissipation} reads
\begin{equation*}
\partial_t \gr(x,t)  = \partial^2_x  \gr(x,t)  - V' \big( \gr(x,t) \big) \, ,
\end{equation*}
with the potential \eqref{eq: potential}
\begin{equation}
\label{eq: rate transition V}
V \big( \gr \big) =  -  (1 - \gga)^2 (1 -\gr ) \gr  +  2 \gga^2 (1 - \gr)^2 \gr^2      \, .
\end{equation}
For $\gga < 1/2$, the potential $V$ has a unique minimum, instead $V$ is a double well when $\gga > 1/2$.
In the latter case a phase transition may occur as there might be several solutions of \eqref{eq: steady}
which can be stable or not with respect to the fluctuating hydrodynamic. 
Depending on the value of the parameters and of the boundary conditions, the steady can be concentrated on more than one profile, leading therefore to a phase transition. 
Some consequences of these transitions on the current large deviations will be examined in section 
\ref{sec: phase transition}.

\section{Large deviations}

\subsection{The functional}

We turn now to the large deviations of the currents and density. The corresponding functional can be completely characterized at the macroscopic level in terms of the coefficients $\gs,D$ (\ref{eq: sigma}-\ref{eq: D}) and $A,C$ \eqref{eq: AC}.

\medskip

On average, the density follows the hydrodynamic equation \eqref{eq: dissipation} and the 
macroscopic currents $\{ \dQ(x,t),\dK(x,t) \}$ obey \eqref{eq: mean conservative}, \eqref{eq: K macro}.
The probability of observing the evolution of an atypical density
profile $\gr(x,t)$ and atypical macroscopic currents $\{ \dQ(x,t),\dK(x,t) \}$
for $0 < t < T$ is given for large $N$ by 
\begin{equation}
\bbP_{[0,T]}  \left( \big\{ \gr, \dQ,\dK \big\} \right) \sim \exp \left[ - 
N \, \cI_{[0,T]} (\gr, \dQ,\dK) \right]
\label{eq: LDF}
\end{equation}
where  $\cI_{[0,T]}= \infty$ if the evolution does not satisfies the conservation law \eqref{eq: macro conservation} and the boundary conditions \eqref{eq: boundary}.
Otherwise, $\cI_{[0,T]}$ is defined by
\begin{eqnarray}
\label{eq: functional}
\cI_{[0,T]}( \gr, \dQ,\dK)
= \int_0^T  dt \int_0^1 dx \;
\left\{ \frac{\big( \dQ(x,t) + D\big(\gr(x,t) \big) \partial_x \gr (x,t) \big)^2}{2 \gs \big( \gr (x,t)\big)}
+ \Phi \Big( \gr(x,t) , \dK(x,t) \Big) \right\} \, , \nonumber \\
\end{eqnarray} 
with 
\begin{eqnarray}
\label{eq: dissipative}
\Phi (\gr, \dK) = C(\gr) + A(\gr) - \sqrt{\dK^2 + 4  A(\gr) C(\gr)}
+ \dK \log \left( \frac{\sqrt{\dK^2 + 4  A(\gr) C(\gr)} + \dK}{2 C(\gr)} \right) \, .
\end{eqnarray} 
If $C(\gr)=0$, then $\Phi$ becomes 
\begin{eqnarray*}
\Phi (\gr, \dK) = 
\begin{cases}
A(\gr) + \dK - \dK \log \left( \frac{- \dK }{ A(\gr)}\right) , \qquad  & {\rm if} \quad  \dK \leq 0,\\
\infty, \qquad  & {\rm if} \quad  \dK >0 \, .
\end{cases}
\end{eqnarray*} 
This can be understood from \eqref{eq: dissipative} by taking the limit $C(\gr) \to 0$ and taking into account the fact that 
for $C(\gr)=0$, there is no creation so that $\dK \leq 0$.
A symmetric expression holds for $A(\gr) =0$.

Note that $\cI_{[0,T]}( \gr, \dQ,\dK) = 0$ for typical evolutions, i.e. when \eqref{eq: macro conservation}, \eqref{eq: mean conservative} and \eqref{eq: K macro} are satisfied.

\medskip

In higher dimensions, a similar expression for the large deviation functional holds. The coefficients $D$ and $\gs$ should be replaced by matrices and $\dQ$ becomes a vector field (see \cite{BDGJL4,BDL}), but the non-conservative part is unchanged.

\subsection{Heuristic derivation of the large deviations}

We turn now to a heuristic derivation of the large deviation principle \eqref{eq: LDF}. 
In the functional $\cI_{[0,T]}$, the contribution of the two currents $\dQ(x,t),\dK(x,t)$ splits into two independent parts.

The first term of the functional involves only the conservative part of the current and it can be interpreted as the local gaussian fluctuations of $\dQ(x,t)$ around its mean $- D\big(\gr(x,t) \big) \partial_x  \gr (x,t)$ (see \eqref{eq: mean conservative}) with a variance $\gs \big(\gr(x,t) \big)$ \eqref{eq: sigma}.  Similar expressions have been obtained in \cite{BD1,BD2,BDGJL4} for diffusive type dynamics.

The contribution $\Phi$ of the non-conservative current can be understood as follows. 
During the time interval $[t,t+dt]$, the density $\gr(x,t)$ remains essentially constant in the small region $[x,x+dx]$. During $[t,t+dt]$, particles are created in $[x,x+dx]$ according to a process
$k^+(x)$ which is the sum of the independent microscopic processes 
at each site in the interval $\{Nx, N(x + dx)\}$.
 As the density remains essentially constant and the local equilibrium is maintained, the process $k^+(x)$ is  the sum of $N dx$ independent Poisson processes with rate 
$\sum_i c(i,\eta_i) (1 -\eta_i) \simeq \bbC N dx $, where $\bbC= C(\gr(x,t))$ \eqref {eq: AC}.
Thus $k^+(x)$ is a Poisson process with a large deviation function given for large $N$ by 
\begin{eqnarray*}
P \left( \int_0^{dt}  k^+ (x)  = \gga^+ N dx dt \right) \simeq
\exp \left( - N dx dt \Psi_\bbC (\gga^+)  \right)
\end{eqnarray*}
where $\int_0^{dt}  k^+ (x)$ stands for the number of particles created in $[x,x+dx]$ during the time interval $[t,t+dt]$ and 
\begin{eqnarray}
\label{eq: Poisson}
\Psi_\bbC (\gga^+) = \gga^+ \,  \log \left( \frac{\gga^+}{ \bbC}\right) - \gga^+ + \bbC \, .
\end{eqnarray}

In the same way, particles are annihilated in $[x,x+dx]$ according to a Poisson process $k^-(x)$ with rate $\bbA N dx = A(\gr(x,t)) N dx$. 
When $N$ tends to infinity, the large deviations of $k^-(x)$ are given by the functional $\Psi_\bbA$ \eqref{eq: Poisson}.
Thus observing a large deviation $\dK(x,t) = k$ of the non-conservative current in $[x,x+dx]$ during the time interval $[t,t+dt]$
 boils down to optimize the large deviations of both processes $k^+ (x),k^-(x)$ under the constraint that 
$k^+(x) - k^-(x) = k N dx dt$ 
\begin{eqnarray}
\label{eq: Poisson LD}
\Phi (\gr(x,t), k) = 
\inf_{\gga^+,\gga^- \atop \gga^+ - \gga^- = k} \left\{  \Psi_\bbC (\gga^+)  + \Psi_\bbA (\gga^-) 
\right\} 
= \inf_{\gga} \left\{  \Psi_\bbC (\gga)  + \Psi_\bbA  (\gga - k) 
\right\} \, .
\end{eqnarray}
Adding the local contributions for each space-time intervals $dx$ and $dt$, we recover the 
non-conservative part of the functional \eqref{eq: functional}.

\medskip

The previous heuristics are based on the conservation of local equilibrium (thanks to the fast stirring) and the summation of all the local fluctuations (Gaussian for the conservative current and Poissonnian for the non-conservative current). A similar {\it additivity principle} has been already used in the framework of diffusive systems \cite{BD1,BD2}.
A rigorous mathematical proof of the large deviations \eqref{eq: LDF} can be found in \cite{BL} for the microscopic model introduced in section \ref{subsec: microscopic}, i.e. $D(\gr) =1$ and $\gs(\gr)= 2 \gr(1-\gr)$ .

\subsection{Density  large deviations}

The large deviations for the density
have been considered in \cite{JLV} and these results can be recovered from the functional \eqref{eq: functional}. The asymptotic probability of observing an atypical evolution $\gr$ for $0 < t < T$ is given for large $N$ by 
\begin{equation}
\bbP_{[0,T]}  \left(  \gr  \right) \sim \exp \left[ - 
N \, \cG_{[0,T]} (\gr) \right]
\ {\rm with} \quad
\cG_{[0,T]} (\gr) = \inf_{\dQ,\dK} \cI_{[0,T]} (\gr, \dQ,\dK)
\label{eq: LDF densite}
\end{equation}
where  the functional $\cI_{[0,T]}$ is optimized over the currents $\{ \dQ,\dK \}$ satisfying the conservation law \eqref{eq: macro conservation}.

It is convenient to rewrite the conservative current in terms of a chemical potential $H$
\begin{eqnarray}
\label{eq: def H}
\dQ(x,t) = - D (\gr(x,t))  \partial_x \gr(x,t) + \gs\big( \gr(x,t) \big) \partial_x H(x,t)
\end{eqnarray}
where $H$ has boundary condition $H(0,t)= 0$.
Thus \eqref{eq: functional} becomes
\begin{eqnarray*}
\cI_{[0,T]}( \gr, \dQ,\dK)
= \int_0^T  dt \int_0^1 dx \;
\left\{ \frac{ \gs \big( \gr (x,t)\big)}{2} \big(  \partial_x H(x,t) \big)^2
+ \Phi \Big( \gr(x,t) , \dK(x,t) \Big) \right\} \, .
\end{eqnarray*} 
Optimizing this functional over $H,\dK$ under the constraint of the conservation law \eqref{eq: macro conservation} implies $ H(1,t) =0$ at any time and  
\begin{eqnarray}
\label{eq: temp}
H(x,t) = - \partial_2 \Phi \Big( \gr(x,t) , \dK(x,t) \Big) \, ,
\end{eqnarray} 
where $\partial_2 $ stands for the derivative wrt $\dK$.
From \eqref{eq: Poisson LD} and \eqref{eq: temp}, we see that  there is $\gga$ such that
$H(x,t) = \log \left( {\gga \over C(\gr(x,t))} \right) = - \log \left( {\gga - \dK(x,t) \over A(\gr(x,t))} \right)$.
This implies that
\begin{eqnarray}
\label{eq: temp 1}
\dK(x,t) = C(\gr(x,t)) \exp \big( H(x,t) \big) - A(\gr(x,t)) \exp \big(- H(x,t) \big)
\end{eqnarray} 
and thus we recover the expression of  \cite{JLV}
\begin{eqnarray*}
&& \cG_{[0,T]}( \gr) =  \int_0^T  \int_0^1 dt \,  dx \;
\frac{ \gs \big( \gr \big)}{2} \big(  \partial_x H \big)^2
+ A( \gr) \Big( 1- \exp(-H) (H + 1)  \Big)  \\ 
&& \qquad \qquad \qquad \qquad \qquad  
+ C( \gr) \Big( \exp(H) (H - 1) -1  \Big) \, ,
\end{eqnarray*} 
where  $H$ is defined such that the following equation holds
\begin{eqnarray*}
\partial_t \rho =  \partial_x \Big( D (\gr)  \partial_x \gr - \gs\big( \gr \big) \partial_x H \Big)
+ C(\gr) \exp \big( H \big) - A(\gr) \exp \big(- H \big) \, .
\end{eqnarray*}
The previous equation is obtained from the conservation law \eqref{eq: macro conservation}
combined with \eqref{eq: def H} and \eqref{eq: temp 1}.


\section{Consequences of the large deviations}
\label{sec: consequences}

In this section, we focus on new the features of the current large deviations which do not exist for purely conservative dynamics.

\subsection{Local variance}
\label{sec: variance}

In practice large deviations are "rarely" observed, however one can often understand the fluctuations close to a steady state by expanding the large deviation functional. 
Using this approach, the cumulants of the current were predicted in \cite{BD1} for general diffusive dynamics and the results were confirmed numerically for some specific models in \cite{BD3,HG}.
We follow the same path and focus on the second order expansion which is related to the variance of the current (when there is no phase transition).

\begin{figure}[h]
\begin{center}
\leavevmode
\epsfbox{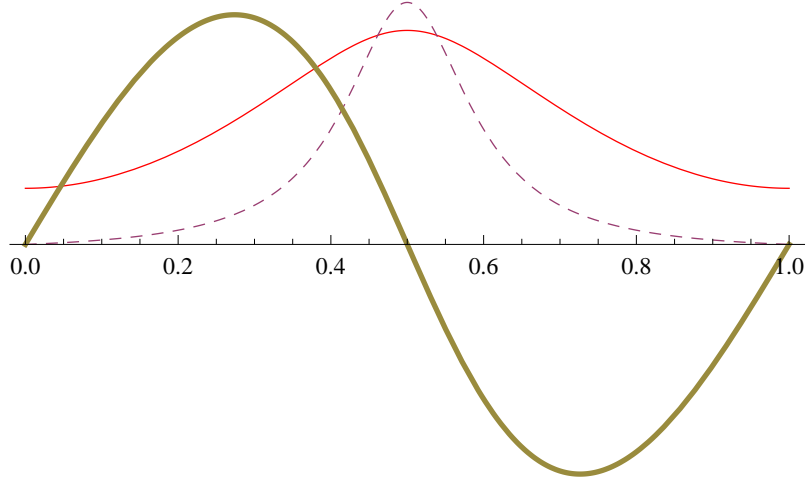}
\caption{For a  function $\gl(x)$ (dashed line), the optimal deviation $f(x)$ (thick line) from the steady $\bar \gr$
 and the optimal current $q(x)$ (thin line) have been rescaled to be represented on the same graph (the vertical axis unit is irrelevant). The function $\gl(x)$ is peaked around $x=1/2$ in order to impose a maximum current $q$ in the region  around $1/2$. The density adapts so that its gradient facilitates a larger flux in the region around $1/2$. Particles are created to the left of $1/2$ and removed on the right.} 
\label{fig: profile courant}
\end{center}
\end{figure}

As the number of particles is non-conserved, the time integrated currents are not constant through the system and the deviations  of the current depend on  where it is measured (this was not the case for the conservative dynamics). In order to illustrate this fact, we are going to compute the deviations of the total conservative current and of  a local conservative current (measured in a smaller region of the bulk). The mechanisms at play are different. In particular, to increase the current from left to right in a given region, extra particles are created to the left of this region and depleted to its right (see figure \ref{fig: profile courant}).
This is reminiscent of the vortices which induce local current fluctuations in two-dimensional models \cite{BDL}.

\medskip

Given a function $\gl (x)$ with $x \in [0,1]$, we are interested in the large deviations of the integrated current 
$ \int_0^1  dx \,   \gl(x) Q(x,T) = \int_0^T dt \, \int_0^1  dx \,   \gl(x) \dQ(x,t)$ over a very long time interval $[0,T]$
\begin{equation*}
\cF (\gl) = \lim_{T \to \infty} \lim_{N \to \infty} \; \frac{1}{NT} 
\log \bbE_{[0,T]}  \left( \exp \left(   \int_0^1  dx \,   \gl(x) Q(x,T) \right) \right) \, . 
\end{equation*}
This boils down to minimizing the time dependent large deviation functional \eqref{eq: functional}
\begin{equation}
\cF (\gl) =  \lim_{T \to \infty} \; \frac{1}{T} \sup_{\gr, \dQ,\dK} \left\{ \int_0^1 \, dx \, \gl(x) \dQ(x,T) -  \cI_{[0,T]}( \gr, \dQ,\dK) \right\} \, . 
\label{eq: LD legendre}
\end{equation}
For some dynamics \cite{BD2,BDGJL4}, this is equivalent to a stationary variational problem and the supremum can be taken on time independent profiles  $(\gr,\dQ,\dK)$.  
In this case, the conservation law \eqref{eq: macro conservation} implies $\dK(x) = \partial_x \dQ(x)$ and the large deviation functional \eqref{eq: functional} reduces to 
\begin{eqnarray}
\hat \cI( \gr, \dQ)
= \int_0^1 dx \;
\left\{ \frac{\big( \dQ(x) + D\big(\gr(x) \big) \partial_x \gr (x) \big)^2}{2 \gs \big( \gr (x)\big)}
+ \Phi \Big( \gr(x) , \partial_x \dQ(x) \Big) \right\} \, .
\label{eq: hat I}
\end{eqnarray} 
It is then enough to consider the simpler variational problem
\begin{eqnarray}
\label{eq: lagrange func}
\cF( \gl) = \sup_{\gr, \dQ} \left\{ \int_0^1 \, dx \, \gl(x) \dQ(x) -  \hat \cI( \gr, \dQ) \right\} \, .
\end{eqnarray}

It is in general not true that the functional \eqref{eq: LD legendre}
can be reduced to \eqref{eq: lagrange func} for any microscopic dynamics, i.e. for any functions $D,\gs$, $A,C$ and $\gl$. 
Some counter-examples leading to dynamical phase transitions can be found in the case of conservative dynamics in 
\cite{BD2,BDGJL4}. 
In section \ref{sec: phase transition}, we will construct a new example of phase transition in the non-conservative case.
When the potential $V$ \eqref{eq: potential} is convex, we expect that \eqref{eq: lagrange func}
gives the correct expression for small deviations close to the steady state current. Indeed in this case the steady state $\bar \rho$ \eqref{eq: steady} is unique and stable with respect to the hydrodynamic evolution \eqref{eq: dissipation}. Thus a small perturbation of the current should not lead to the bifurcations observed in the  dynamical phase transitions \cite{BD3}.

\begin{rem}
The previous conjecture on the equality between \eqref{eq: LD legendre}
and \eqref{eq: lagrange func}
can be derived when $D$ is a constant, $\gs$ is concave, $A(\gr)=a \gr$ and $C(\gr)=c(1-\gr)$ are linear (and thus $V$ is quadratic). 
Recall that for the SSEP, one has $D=1$, $\gs (\gr) = 2 \gr (1-\gr)$.
We follow the argument in \cite{BDGJL4} devised for the SSEP.
First note that  the functional $\cI_{[0,T]}$ can be rewritten (see e.g. \cite{BL}) as
\begin{equation*}
\cI_{[0,T]} \left(\rho,\dQ,\dK\right)  
= \int_0^T dt \, \Psi (\rho_t, \dQ_t)+ \Phi (\rho_t , \dK_t) \, ,
\end{equation*} 
where we set for any functions $f(x),q(x),k(x)$ with $x$ in $[0,1]$
\begin{eqnarray*}
\Psi (f,q) = \sup_{H} \; \Psi_H (f,q) =
\sup_{H} \;  
\left\{
\int_0^1 dx \; \big( q (x) + \partial_x f(x) \big) H(x)
 -\frac{1}{2} \sigma(f(x))H(x)^2 
\right\} \, ,
\end{eqnarray*}
and
\begin{eqnarray*}
\Phi(f,k) = \sup_G \;  \Phi_G (f,k)
= 
\sup_G \; 
\left\{ 
\int_0^1 dx \;  k(x)  G(x) - C(f(x))(e^{G(x)}-1) -  A(f(x))(e^{-G(x)}-1)
\right\}
\, .
\end{eqnarray*}
Note that the representation holds for any $A$ and $C$.

For any $H$, the functional $\Psi_H (f,q)$ is convex jointly in $(f,q)$ because $\gs$ is concave. In the same way, for any $G$, the functional $\Phi_G(f,k)$ is jointly convex in $(f,k)$ since $A$ and $C$ are linear. As $\Psi$ and $\Phi$ are defined as a supremum of convex functions, they are also convex.
Thus for any trajectory $(\rho,\dQ,\dK)$
\begin{equation*}
\Psi \left( \frac{1}{T} \int_0^T dt \,  \rho_t, \frac{1}{T} \int_0^T dt \,  \dQ_t \right)
+
\Phi \left(\frac{1}{T} \int_0^T dt \, \rho_t , \frac{1}{T} \int_0^T dt \,  \dK_t \right) 
\leq 
\frac{1}{T} \cI_{[0,T]} \left(\rho,\dQ,\dK\right)  \, .
\end{equation*} 
This inequality implies that the minimum in \eqref{eq: LD legendre} can be reduced to time independent profiles.
Indeed, any time dependent evolution $(\rho,\dQ,\dK)$ which produces a total current $q = \frac{1}{T} \int_0^T dt \,  \dQ_t $ has a large deviation cost greater or equal than the time independent profiles obtained by averaging this trajectory over time.   
\end{rem}

\medskip

Given the function $\gl$, we are going to expand $\cF(\gep \gl)$ (given in \eqref{eq: lagrange func})
to the second order in $\gep$.
For a small shift of the current, we expect that the minimum is located close to 
the steady state $\bar \gr$ defined in \eqref{eq: steady}.
We introduce the perturbation 
\begin{eqnarray*}
\gr(x) = \bar \gr(x) + \gep {f(x) \over D\big(\bar \gr(x) \big) }, \qquad 
\dQ(x) = - D\big(\bar \gr(x) \big) \partial_x \bar \gr (x) + \gep q(x) \, ,
\end{eqnarray*} 
with $f(0) =f(1) =0$ to satisfy the boundary conditions \eqref{eq: boundary}.

Expanding the functional $\hat \cI$ \eqref{eq: hat I} to the second order in $\gep$ leads to
\begin{eqnarray}
\label{eq: functional stat}
&& \cF(\gep \gl) = -  \gep \int_0^1 dx \; \gl(x) \, D\big(\bar \gr(x) \big) \partial_x \bar \gr (x)\\
&& \qquad \qquad + \gep^2 
\inf_{q,f} \left\{ \int_0^1 dx \; \gl(x) q(x) 
- \left ( \frac{\big( q(x) + f' (x)  \big)^2}{2 \gs \big( \bar \gr (x)\big)}
+ \frac{ \big( q'(x) + U(x) \, f(x)  \big)^2}{2 (A({\bar \rho}(x))+C({\bar \rho}(x)))}  \right)
\right\} \, , \nonumber
\end{eqnarray}
where $U(x)={A'({\bar \rho}(x)) - C'({\bar \rho}(x)) \over D({\bar \rho}(x))} \geq 0$ by the convexity assumption of $V$.
We introduce 
\begin{eqnarray}
\label{eq: changement variables}
\gp (x) = \frac{ q(x) +  f' (x)}{\gs ( \bar \gr (x))}, \qquad 
\psi (x) =  \frac{ q'(x) + U(x) \, f(x)}{A({\bar \rho}(x))+C({\bar \rho}(x))} \, , 
\end{eqnarray}
and optimize \eqref{eq: functional stat} over $f$ and $q$.
This implies
\begin{eqnarray*}
\begin{cases}
\gl(x) = \gp(x) - \partial_x  \psi (x)  \\
0 = - \partial_x \gp(x) + U(x) \psi(x) 
\end{cases}
\end{eqnarray*}
with the boundary conditions $\psi(0)=\psi(1) =0$ and $f(0)= f(1) =0$.
We deduce that 
\begin{eqnarray}
\label{eq: solutions}
\begin{cases}
\gl' (x) = U(x) \psi(x) - \partial_x^2 \left( \psi(x) \right) \\
\gp(x) = \gl(x) + \partial_x  \psi (x) \\
\psi(0)=\psi(1) = f(0)= f(1) =0 
\end{cases}
\end{eqnarray}
Since $U(x) \geq 0$, the solution $(\gp,\psi)$ of \eqref{eq: solutions} is unique.
The current and density profiles $(q,f)$ can then be determined uniquely since the boundary are fixed 
$q'(0)=q'(1) =0$ and $f(0)= f(1) =0$.

\bigskip

\noindent
$\bullet$ {\it Total current deviations.}

The variance of the total current $ \int_0^1 dx \,  q(x)$ can be obtained with a constant Lagrange parameter
$\gl(x) = \gl$. In this case, the solutions of \eqref{eq: solutions} are $\gp(x) = \gl$ and $\psi(x) =0$. Plugging this in \eqref{eq: functional stat} leads to
\begin{eqnarray}
\label{eq: courant total variance}
\cF(\gep \gl) = -  (\gep \gl) \int_0^1 dx  \, D\big(\bar \gr(x) \big) \partial_x \bar \gr (x)
+ \frac{(\gep \gl)^2}{2} \int_0^1 dx  \, \gs \big(\bar \gr(x) \big) \, .
\end{eqnarray}
The variance of the total conservative current should be given by $\int_0^1 dx  \, \gs \big(\bar \gr(x) \big)$.

The result is the same as the one obtained in the purely conservative case \cite{BD1}, however it is important to 
note that the minimizers of the variational problem \eqref{eq: functional stat} can be different from the conservative case and that the optimal current is, in general,  no longer uniform through the system.

\bigskip

\noindent
$\bullet$ {\it Local current deviations.}

In order to solve \eqref{eq: solutions} for local current fluctuations, we consider the simple case 
where $\bar \gr$ is constant so that all the differential equations have constant coefficients (denoted by $A,C,U,\gs$).
For example, this is the case if the reservoirs \eqref{eq: boundary}  have equal densities $\gr_a = \gr_b = \bar \gr$ 
and $V'(\bar \gr) = 0$.
We rewrite  \eqref{eq: changement variables} and \eqref{eq: solutions}
\begin{eqnarray}
\label{eq: solutions 2}
\begin{cases}
\gl' (x) = U \psi(x) - \partial_x^2 \left( \psi(x) \right), \qquad \gp(x) = \gl(x) + \partial_x  \psi (x) \, , \\
 q(x) +  f' (x) = \gs \gp (x), \qquad 
q'(x) + U \, f(x) = (A + C) \psi (x) \, ,\\
\psi(0)=\psi(1) = f(0)= f(1) =0 \, .
\end{cases}
\end{eqnarray}
Recall that by the convexity of $V$ then $U \geq 0$.
The solutions are given by 
\begin{eqnarray}
\begin{cases}
\psi(x) = a_0 \sinh(\sqrt{U} x) - \int_0^x \, dy \, {\gl'(y) \over \sqrt{U}} 
\sinh \big( \sqrt{U} (x-y) \big) \, ,\\
\gp(x) = \gl(x) + a_0 \sqrt{U} \cosh (\sqrt{U} x) - \int_0^x \, dy \, \gl'(y) \cosh \big( \sqrt{U} (x-y) \big) \, ,
\end{cases}
\end{eqnarray}
with $a_0 = { \int_0^1 \, dy \, {\gl'(y) \over \sqrt{U}} 
\sinh \big( \sqrt{U} (1-y) \big) \over \sinh(\sqrt{U}) }$.
Since
\begin{eqnarray*}
\begin{cases}
q''(x) - U q = - U \gs \gp (x) + (A + C) \psi' (x)=   (A + C - U \gs) \gp (x) - (A + C)  \gl(x) , \\
f'' (x) - U \, f(x) = \gs \gp'(x) - ( A + C) \psi (x)  =
(U \gs - A - C) \psi (x) \, ,\\
q' (0)= q' (1) = f(0)= f(1) =0 \, .
\end{cases}
\end{eqnarray*}
Finally the optimal current and density are given by
\begin{eqnarray}
\begin{cases}
q(x) = a_1  \cosh (\sqrt{U} x) + \int_0^x \, dy \, {(A + C - U \gs) \gp (y) - (A + C)  \gl(y) \over \sqrt{U}} 
\sinh \big( \sqrt{U} (x-y) \big) \, ,\\
f(x) = a_2 \sinh(\sqrt{U} x)  + (U \gs - A - C) \int_0^x \, dy \,  {\psi (y) \over \sqrt{U}} 
\sinh \big( \sqrt{U} (x-y) \big) \, ,
\end{cases}
\end{eqnarray}
with $a_1,a_2$ such that the boundary conditions are satisfied.

When $\gl$ is a constant, the optimal density and current associated to the deviations of the total current are constant: $f=0$ and $q = \gl \gs$. However to achieve a local current deviation, the interplay between the stirring process and the Glauber dynamics is more complex and the density is no longer uniform as depicted in figure \ref{fig: profile courant}.

\subsection{Phase transition}
\label{sec: phase transition}

In this section, we consider a slightly different dynamics in order to explain in the simplest framework a mechanism of phase transition for the current large deviations.
The system is on a periodic chain $\{1,N\}$ without reservoirs.
Instead of the SSEP, the stirring process is now the Weakly Asymmetric Exclusion Process (WASEP) with jump rates to the right given by $N^2 (1 + \frac{\nu}{ N}) dt$ and to the left is given by $N^2 (1 - \frac{\nu}{ N}) dt$. The drift is of order $\frac{\nu}{N}$ and $\nu$ is a fixed parameter.
This stirring process is combined to the Glauber dynamics with rates  \eqref{eq: rate transition}.
For $\gga >1/2$, this leads to a double well potential $V$ \eqref{eq: rate transition V}  
with minima $\gr^+$, $\gr^- = 1 - \gr^+$ and a local maximum at $1/2$. 

The currents are defined as in section \ref{sec: hydrodynamic limit}. 
The macroscopic equation \eqref{eq: mean conservative} for the conservative current becomes
\begin{eqnarray}
\label{eq: current drift}
\dQ(x,t)  = - \partial_x \gr(x,t) + \nu \gs \big( \gr(x,t) \big) \, ,
\end{eqnarray}
with  $\gs(u) = 2 u (1-u)$.
Thus the new hydrodynamic evolution can be written
\begin{equation}
\partial_t \gr(x,t)  = \Delta \gr(x,t)  - \nu \partial_x \big(\gs(\gr(x,t)) \big) - V' \big( \gr(x,t) \big)  \, .
\label{eq: new hydro}
\end{equation}

\medskip

The  stationary measure of this dynamics exhibits a phase transition as the steady state concentrates on two different density profiles in the thermodynamic limit $N \to \infty$.
To see this, we first notice that the constant profiles equal to $\gr^+, \gr^-$ and $1/2$ are all stationary solutions of the hydrodynamic equation \eqref{eq: new hydro}. 
To guess the structure of the stationary state, one has to consider the stochastic corrections provided
by the fluctuating hydrodynamics \cite{DFL,OS}.
A  noise term depending non-linearly on the local density has to be added to \eqref{eq: new hydro} (see 
\cite{DFL} for an exact expression).
Thus only $\gr^+, \gr^-$ are stable with respect to the noisy equation and $1/2$ is not. Depending on the values of $\gga,\nu$ there can be also other stationary solutions of \eqref{eq: steady} which are not constant, but they will be unstable.
Thus we say that  the stationary measure has a phase transition: the steady state concentrates in the large $N$ limit on the two constant profiles $\gr^+, \gr^-$ with equal probability due to the symmetry of the rates \eqref{eq: rate transition}.
(The previous argument is mainly heuristic and we postpone a mathematical justification of the  occurrence of a phase transition to a futur work.)

\medskip

We will see now that the phase transition of  the stationary measure has also implications on the current large deviations.
Adding a weak drift $\nu$ to the stirring process modifies only the cost of the conservative current deviations in the  large deviation functional \eqref{eq: functional}.
The new functional reads
\begin{eqnarray}
\label{eq: functional drift}
\cI^\nu_{[0,T]} ( \gr, \dQ,\dK)
= \int_0^T  dt \int_0^1 dx \,
\left\{ \frac{\big( \dQ(x,t) + \partial_x \gr (x,t) - \nu \gs(\gr (x,t)) \big)^2}{2 \gs \big( \gr (x,t)\big)}
+ \Phi \Big( \gr(x,t) , \dK(x,t) \Big) \right\} \, , \nonumber \\
\end{eqnarray} 
where $\Phi$ is still given by \eqref{eq: dissipative}, the conductivity $\gs(u) = 2 u(1-u)$ and the diffusion coefficient $D$ is now equal to $1$. We refer to \cite{BD2,BD3,BDGJL4} for a study of the large deviations in conservative dynamics with weak drifts.

\begin{figure}[h]
\begin{center}
\leavevmode
\psfragscanon 
\psfrag{q}[t][l]{{\small $\nu \gs(\gr^+)$}}
\psfrag{r}[t][c]{{\small $\nu/4$}}
\psfrag{G}[r][l]{{\small $G(q)$}}
\psfrag{T}[t][l]{{\small $T$}}
\psfrag{t}[t][c]{{\small $\ga T$}}
\psfrag{R}[l][b]{{\small $\gr(x,t)$}}
\centerline{
\psfig{figure=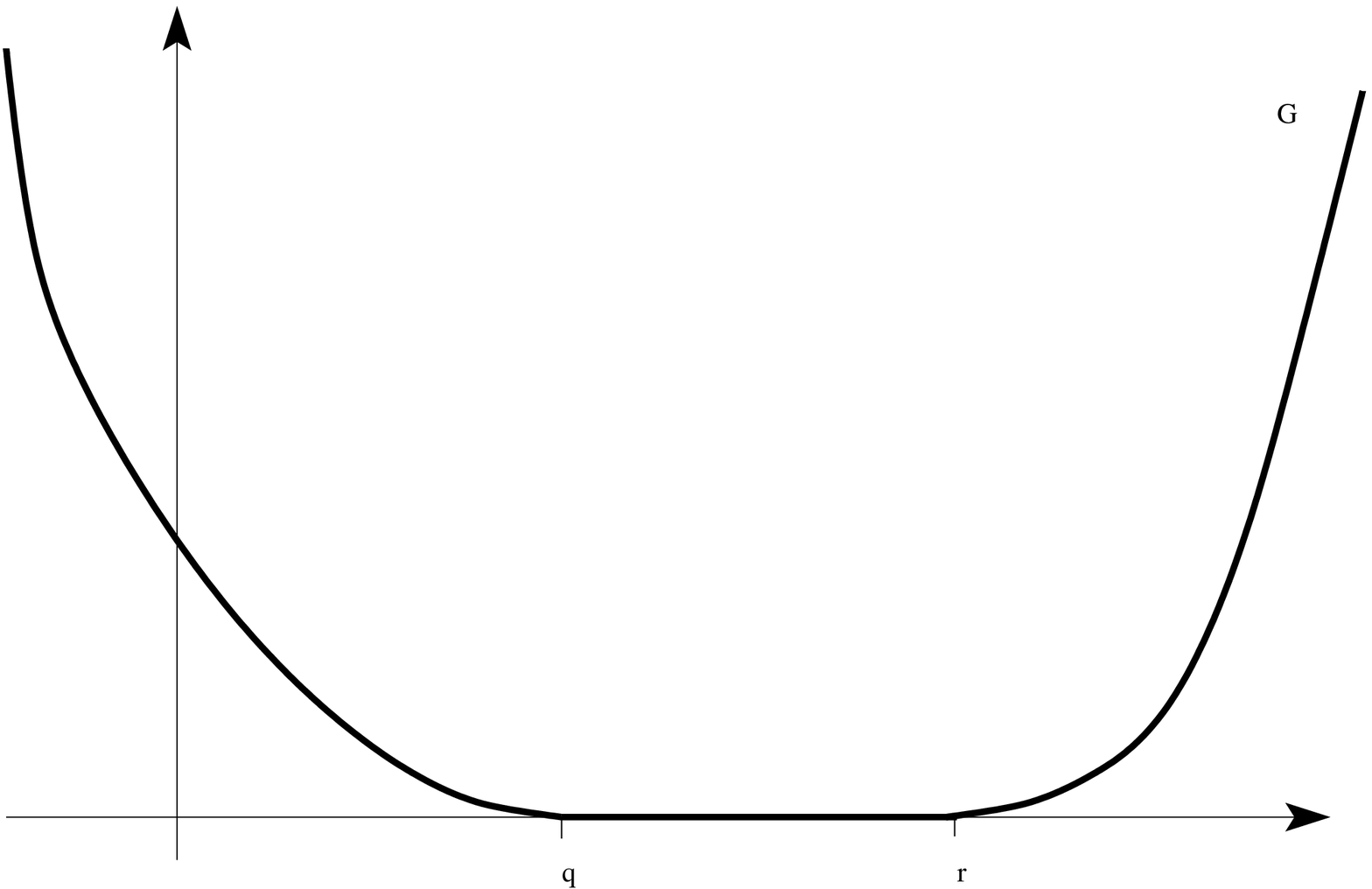,height=35mm}
\hspace{1cm}
\psfig{figure=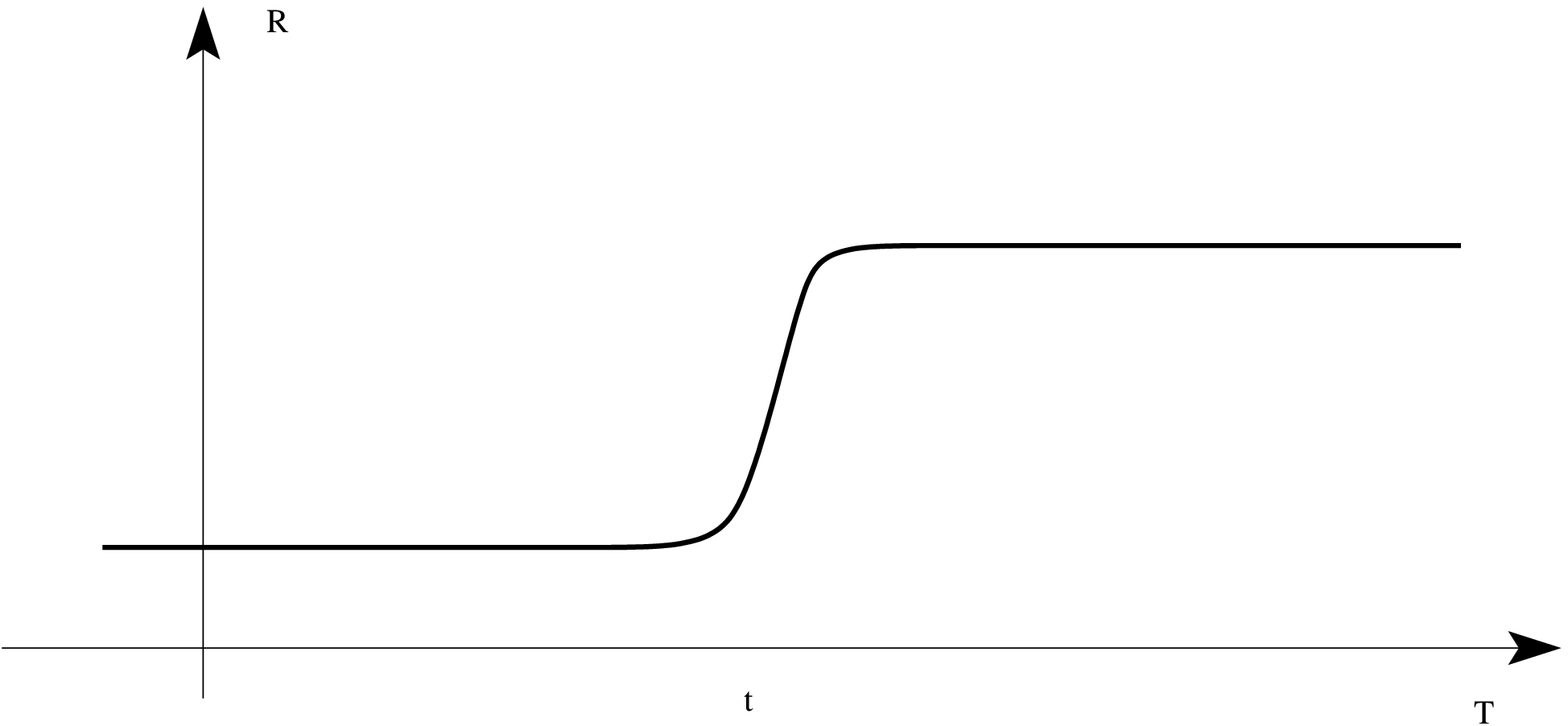,height=35mm}}
\caption{The current large deviation function $G$ is represented on the left.
On the right, a switch at time $\ga T$ between the densities $\gr^+$ and $1/2$ is depicted.} 
\label{fig: transition}
\end{center}
\end{figure}

As the steady states concentrates on $\gr^+,\gr^-$, the mean conservative current
\eqref{eq: current drift} is $\nu \gs(\gr^+) = \nu \gs (\gr^-)$ (in the large $N$ limit). We are going to compute the cost of an increase of the total current over a very large time interval. For any shift of the current $q \in [  \nu \gs(\gr^+), \nu /4]$
\begin{equation}
\lim_{T \to \infty} \lim_{N \to \infty} \; \frac{1}{T N} \, \log
\bbP_{[0,T]}  \left( \frac{1}{T} \int_0^1 dx \, Q(x,T) \simeq q \right) = 0 \, .
\label{eq: zero LD}
\end{equation}
The strategy to minimize the functional \eqref{eq: functional drift} goes as follow.
To produce a current  close to $\nu T /4 = \nu \gs(1/2)T$ over a very long time $T$, the system (which originally starts close to $\gr^+$ or $\gr^-$) can evolve to the constant profile $1/2$ within a time of order 1. This first move has a cost exponential in $N$ but independent of $T$ so that it is irrelevant in the limit \eqref{eq: zero LD}.
Then the density remains equal to $1/2$ for the rest of the time and therefore the system produces a conservative current \eqref{eq: current drift} asymptotically equal to $\nu T /4$  for large $T$.
As $1/2$ is a metastable profile this has no large deviation cost (at least in the scaling considered in \eqref{eq: zero LD}). For an intermediate current $q = \ga   \nu \gs(\gr^+) + (1-\ga) \nu /4$ (with $\ga \in ]0,1[$), the deviation can be achieved by a time-dependent density profile which is close to $\gr^\pm$ during a time $\ga T$ and 
then close to $1/2$ during a time  $(1-\ga) T$. 
The optimal cost remains equal to 0, but the time dependent variational principle 
\eqref{eq: LD legendre} cannot be reduced to \eqref{eq: lagrange func}
even for arbitrarily small current deviations.
A flat piece in the large deviation function (see Figure \ref{fig: transition}) is a sign of a phase transition.

\subsection{The fluctuation relation}
\label{sec: The fluctuation relation}

The large deviation function of the entropy production obeys a very general symmetry property \cite{GC,ECM} which is known to hold for a wide class of stochastic dynamics \cite{K,LS,M}. 
When applied to conservative dynamics in contact with two heat baths, the entropy production can be reinterpreted as the current flowing through the system and therefore a similar symmetry holds for the large deviation function of the current at the microsopic and macroscopic levels \cite{BD1,BDGJL4}.
A generalized notion of detailed balance introduced in \cite{D,BD3} allows to consider an arbitray number of reservoirs and therefore is well suited in the case of bulk creation/dissipation. 
Finally, the deviations of the entropy production for the more realistic case of granular gases was considered in \cite{BGGZ}.

In this section, we recover the fluctuation relation at the macroscopic level by means of the functional \eqref{eq: functional}.

\bigskip

We are going now to investigate the symmetries of the functional  \eqref{eq: functional} with respect to  
time reversal which is a key feature of the Gallavotti-Cohen relation. 
Let $(\gr, \dQ,\dK)$ be an evolution  from time 0 to time $T$ such that the coefficients $A(\gr)$ and $C(\gr)$ are positive.
The functional \eqref{eq: functional} can be rewritten as
\begin{eqnarray}
&&\cI_{[0,T]}( \gr, \dQ,\dK)
= \int_0^T  dt \int_0^1 dx \,
\frac{ \dQ(x,t)^2 + \big( D\big(\gr(x,t) \big) \partial_x \gr (x,t) \big)^2}{2 \gs \big( \gr (x,t)\big)}
+  \frac{ \dQ(x,t)  D\big(\gr(x,t) \big) \partial_x \gr (x,t)}{ \gs \big( \gr (x,t)\big)} \nonumber \\
&& \qquad \qquad \qquad \qquad + \Phi \Big( \gr(x,t) , \dK(x,t) \Big)\, , 
\label{eq: func rev}
\end{eqnarray} 
with $\Phi$ given by  
\begin{eqnarray}
\Phi (\gr, \dK)  &=& C(\gr) + A(\gr) - \sqrt{\dK^2 + 4  A(\gr) C(\gr)}
- \dK \log \left( \frac{\sqrt{\dK^2 + 4  A(\gr) C(\gr)} - \dK}{2 C(\gr)} \right)  \nonumber \\
&& \qquad  + \dK \log \left( \frac{  A(\gr) }{C(\gr)} \right) 
= \Phi (\gr, - \dK) + \dK \log \left( \frac{  A(\gr) }{C(\gr)} \right)\, ,
\label{eq: - phi}
\end{eqnarray} 
where the previous identity follows from \eqref{eq: dissipative} by multiplying the numerator and 
the denominator (in the log) by $\sqrt{\dK^2 + 4  A(\gr) C(\gr)} - \dK$.

Let $S(u)$ be a function such that $S''(u) = 2 \frac{D( u )}{ \gs ( u )}$
(we shall see later on that $S$ can be interpreted as a relative entropy). By integration by parts, one has
\begin{eqnarray*}
&& 2 \int_0^T  dt \int_0^1 dx \,
\frac{ \dQ(x,t)  D\big(\gr(x,t) \big) \partial_x \gr (x,t)}{ \gs \big( \gr (x,t)\big)}
= 
\int_0^T  dt \int_0^1 dx \, \dQ(x,t) \; \partial_x \,\Big[  S' \big( \gr (x,t)\big) \Big]\\
&& \qquad \qquad 
= - \int_0^T  dt \int_0^1 dx \, \partial_x \dQ(x,t)  \, S' \big( \gr (x,t)\big) 
+  \Big( Q(1,T) S' \big( \gr_b \big) - Q(0,T) S' \big( \gr_a \big) \Big) \, ,
\end{eqnarray*}
where  $Q(y,T)=\int_0^T  dt \, \dQ(y,t)$ and  $\gr_a,\gr_b$ are the densities imposed by the left and right reservoirs.
The conservation law \eqref{eq: macro conservation} leads to
\begin{eqnarray*}
&& 2 \int_0^T  dt \int_0^1 dx \,
\frac{ \dQ(x,t)  D\big(\gr(x,t) \big) \partial_x \gr (x,t)}{ \gs \big( \gr (x,t)\big)}\\
&& \qquad 
=  \Big( Q(1,T) S' \big( \gr_b \big) - Q(0,T) S' \big( \gr_a \big) \Big)  +
\int_0^T  dt \int_0^1 dx \, \Big( \partial_t \rho(x,t)- \dK(x,t)  \Big)  \, S' \big( \gr (x,t)\big)\\
&& \qquad 
= \Big( Q(1,T) S' \big( \gr_b \big) - Q(0,T) S' \big( \gr_a \big) \Big) 
- \int_0^T  dt \int_0^1 dx \,  \dK(x,t)\, S' \big( \gr (x,t)\big) \\
&& \qquad \qquad 
+ \int_0^1 dx \, \Big(  S \big( \gr (x,T)\big) - S \big( \gr (x,0)\big) \Big)  \, .
\end{eqnarray*} 
Combining \eqref{eq: func rev}, \eqref{eq: - phi} and the previous identity, one can 
compare the large deviation cost $\cI_{[0,T]}( \gr, \dQ,\dK)$ to the one of its time reversal
$(\hat \gr, \hat \dQ, \hat \dK) = (\gr(T-t), - \dQ(T-t), - \dK(T-t))$
\begin{eqnarray}
\cI_{[0,T]}( \gr, \dQ,\dK) - \cI_{[0,T]}(\hat \gr,\hat \dQ,\hat \dK)  
= \cE_{[0,T]}( \gr, \dQ,\dK)
+ \int_0^1 dx \, \Big(  S \big( \gr (x,T)\big) - S \big( \gr (x,0)\big) \Big) \, .
\label{eq: time reversal}
\end{eqnarray} 
where the {\it entropy production} can be identified as 
\begin{eqnarray}
\label{eq: entropy production}
&& \cE_{[0,T]}( \gr, \dQ,\dK)
= 
\Big( Q(1,T) S' \big( \gr_b \big) - Q(0,T) S' \big( \gr_a \big) \Big)\\
&& \qquad \qquad 
+ 
\int_0^T  dt \int_0^1 dx \, \dK(x,t)\, \Big( -  S' \big( \gr (x,t)\big)
+ \log \left( \frac{  A(\gr(x,t)) }{C(\gr(x,t))} \right)  \Big)  \, . \nonumber  
\end{eqnarray} 
By construction $\cE_{[0,T]}$ is anti-symmetric under time reversal.

We consider dynamics such that $S$ is bounded (which is the case if the (microscopic) number of particle per site is bounded).
Then the large deviation function of the entropy production
\begin{eqnarray*}
\cG (e) 
= \lim_{T \to \infty} 
- \frac{1}{T} \log \; \bbP_{[0,T]} \left( \frac{1}{T} \cE_{[0,T]}( \gr, \dQ,\dK) = e \right)  
\end{eqnarray*} 
satisfies the Gallavotti-Cohen symmetry $\cG (e) = \cG (-e) -e$ (thanks to the identity \eqref{eq: time reversal} and the fact that the term $ \int_0^1 dx \, \Big(  S \big( \gr (x,T)\big) - S \big( \gr (x,0)\big) \Big) $ is bounded uniformly wrt time).

\bigskip

For conservative systems, the entropy production $\cE_{[0,T]}$ \eqref{eq: entropy production} is proportional to the conservative current $Q(0,T)$ flowing through the system and therefore the symmetry holds also for the large deviation functional of the conservative current \cite{BD1,BD3,BDGJL4}. 
For general creation/annihilation bulk rates, the relation between $\cE_{[0,T]}$ and the currents is less straightforward as the density is coupled to the non-conservative current $\dK$.

Simplifications occur when the steering process and the Glauber dynamics are both reversible (locally) with respect to the same Gibbs measure.
Let $\mu_\gr$ be a Gibbs measure with constant density $\gr$ which is invariant for the steering process 
 for any density $\gr$  (see \eqref{eq: dissipation 0}). 
We denote by $\chi(\gr)$ the susceptibility of $\mu_\gr$. 
According to the fluctuation dissipation relation \cite{spohn}, the function $S(u)$ 
satisfies $S''(\gr) = 2 \frac{D(\gr)}{ \gs(\gr)} = \frac{1}{\chi (\gr)}$.
Thus $S'(\gr)$ can be interpreted as the chemical potential conjugated to the density $\gr$ and $S$ as an equilibrium large deviation function.
Suppose that the creation and annihilation are reversible with respect to $\mu_{\gr_0}$ for a given density $\gr_0$,
i.e.
\begin{eqnarray*}
\mu_{\rho_0} \big( c(i,\eta) \eta_i \, \big| \; \{\eta_j\}_{j \not = i}  \big)
= \mu_{\rho_0} \big( c(i,\eta) (1-\eta_i) \, \big| \; \{\eta_j\}_{j \not = i}  \big)
\end{eqnarray*}
where $\mu_{\rho_0} \big( \cdot \, \big| \; \{\eta_j\}_{j \not = i}  \big)$ is the Gibbs measure on $\eta_i$ conditionally to the configuration $\{\eta_j\}_{j \not = i}$ outside the site $i$.
Thus \eqref{eq: AC} can be rewritten
\begin{eqnarray*}
A (\rho) &=& \mu_{\rho} \big( c(i,\eta) \eta_i \big)
= \exp \big( S'(\gr)-  S'(\gr_0) \big) \mu_{\rho} \big( \mu_{\rho_0} \big( c(i,\eta) \eta_i \, \big| \; \{\eta_j\}_{j \not = i}  \big) \big)\\
&=& 
\exp \big( S'(\gr)-  S'(\gr_0) \big) C ( \rho) \, .
\end{eqnarray*}

Using this identity, the entropy production \eqref{eq: entropy production} can be rewritten only in terms of the 
currents
\begin{eqnarray*}
\cE_{[0,T]}( \gr, \dQ,\dK)
= 
 Q(1,T) S' \big( \gr_b \big) - Q(0,T) S' \big( \gr_a \big) 
- S' \big( \gr_0 \big)  \int_0^1 dx \, K(x,T)  \, .
\end{eqnarray*}

\section{Conclusion}

In this paper, we analyzed the current large deviations for non-conservative diffusive dynamics driven out-off equilibrium by external reservoirs. We obtained an expression for the large deviation functional \eqref{eq: LDF} which generalizes the results derived for the conservative dynamics \cite{BDGJL4}-\cite{BD3}.

From the large deviation functional, one can understand how the deviations of the currents and the density are coupled. 
Contrary to the conservative case, there are now two currents instead of one and it is therefore more delicate to derive 
exact expressions of the minimizers of the functional given some constraint (for example given the total currents). 
It would be of interest to derive more systematic predictions for the cumulants of the currents as in \cite{BD1}.

The phase transition for the dynamics considered in this paper is currently not understood at the mathematical level and we plan to study it in the future.

In section \ref{sec: The fluctuation relation}, we showed that the entropy production  \eqref{eq: entropy production}
satisfies the fluctuation relation. One may wonder if other (measurable) physical quantities related to the injected current satisfy similar relations.
In this spirit, 
the exact expression of the large deviation functional \eqref{eq: LDF} could be helpful to solve some controversy on the validity of the fluctuation relation in dissipative systems \cite{FP}.


\end{document}